\documentclass[english,superscriptaddress,pra,twocolumn]{revtex4}
\usepackage[T1]{fontenc}
\usepackage[latin9]{inputenc}
\usepackage{amsmath}
\usepackage{amssymb}
\usepackage{graphicx}

\makeatletter
\@ifundefined{textcolor}{}
{%
 \definecolor{BLACK}{gray}{0}
 \definecolor{WHITE}{gray}{1}
 \definecolor{RED}{rgb}{1,0,0}
 \definecolor{GREEN}{rgb}{0,1,0}
 \definecolor{BLUE}{rgb}{0,0,1}
 \definecolor{CYAN}{cmyk}{1,0,0,0}
 \definecolor{MAGENTA}{cmyk}{0,1,0,0}
 \definecolor{YELLOW}{cmyk}{0,0,1,0}
}

\usepackage{times}

\makeatother

\usepackage{babel}
\begin{document}

\title{Atom-driven multistability in an optomechanical cavity under broken
$\mathcal{PT}$-symmetry}

\author{Hou Ian}

\affiliation{Institute of Applied Physics and Materials Engineering, FST, University
of Macau, Macau}
\begin{abstract}
An optical field inside a Fabry-Perot cavity would exhibit multistability
either when an atomic medium, acting as a classical dielectric, is
filled into the cavity or when the cavity becomes optomechanical where
one reflecting end becomes movable. An external laser is essential
in both cases to drive the cavity mode out of the zero equilibrium
state. We study the equilibrium states of an atom-filled optomechanical
cavity, where the atoms are considered a collective two-level systems
as well as an active component that replaces the role of the driving
laser. Multistability is found when the atomic ensemble is lossy,
for which the complex atom-photon coupling breaks the $\mathcal{PT}$-symmetry
of the system Hamiltonian and matches in a periodic pattern to the
frequencies and the linewidths of the cavity mode and the collective
bosonic mode of the atoms. We show an input-output hystersis cycle
between the atomic mode and the cavity mode to demonstrate the driving
role the atoms replace an extra-cavity laser.
\end{abstract}


\maketitle

\section{Introduction }

When a nonlinear medium is filled inside a Fabre-Perot (FP) cavity
interferometer, an incident laser through the cavity experiences optical
bistability~\cite{gibbs76,Gibbs,bonifacio78}. From a classical point
of view, it is the dual effective round-trip paths inside the nonlinear
medium that permits the incident laser admit two stable values of
transmittivity, the determination of which depends on the intensity
of the laser~\cite{marburger78}. The multi-valued branching of the
transmittivity forms a hystersis cycle typical of the bistability
behavior in the laser input-output power diagram~\cite{gibbs76}.
When the nonlinear medium has Zeeman sublevels that are split by an
optical pumping, the lower branch of the hystersis cycle breaks up
into two, making the system tristable~\cite{kitano81}. In the case
of ring cavity, even multistability is achievable if the filled medium
consists of three-level atoms~\cite{joshi03}.

This kind of optical multistability is not unique to atom-filled FP
cavities. The role played by the nonlinear atomic medium can be substituted
by a movable mirror placed at one end of the cavity~\cite{meystre85,gong09,klimov01}.
The mirror's oscillation deforms the cavity and induces a Kerr-like
nonlinear dispersion, giving rise to a similar hystersis of bistability
induced by the photon-phonon interaction in radiation pressure~\cite{dorsel83}.
Since then, interests in the stability problem is revived~\cite{marquardt06,chang11}
under the context of cavity optomechanics~\cite{marquardt09}. The
mechanical multistability problem has ramified into many branches
of studies, including optomechanical instability~\cite{qian12},
optimal regime for entanglement~\cite{ghobadi11,ian15}, cavity assisted
spin-orbit coupling~\cite{dong11}, and optomechanially induced transparency~\cite{agarwal10,kronwald13,karuza13}.

In this paper, we find out how multistable states arise when both
an atomic medium and an oscillating mirror are present, where the
optomechanical cavity is essentially a quantum tripartite AOM system
composing of the collective bosonic mode of the \emph{atoms}, the
\emph{optical} cavity mode, and the \emph{mechanical} resonator mode
of the mirror. Considering the fact that atoms in cavities can initiate
strong coupling to both the cavity field~\cite{wallquist10} and
the mirror~\cite{hunger10}, strong interfering motions are expected
to appear between the atomic medium and the oscillating mirror. It
has been shown that the mirror can be bistable~\cite{zhang10} or
multistable~\cite{chang11}.

Considering the atoms as two-level systems, our study here, however,
is to test whether the motion of the atoms alone can drive the optomechanical
cavity into multistable equilibria, where the external laser driving
is eliminated, since it is proven that the atoms' excitations can
vibrate like a resonator~\cite{brennecke08} and provide effective
driving and squeezing to the mirror~\cite{ian08}. Presented in Sec.~\ref{sec:Coupling-mechanism}
below, this tripartite coupling model without external driving regards
the motions of the atoms through a collective bosonic mode whose interaction
with the cavity mode is lossy.

As we shall demonstrate in Sec.~\ref{sec:Phase-matching}, it is
found that multistability arises only when this lossy bosonic mode,
which breaks the \emph{$\mathcal{PT}$}-symmetry of the tripartite
system Hamiltonian~\cite{bender98}, has an atom-photon coupling
that matches with its own eigenfrequency and linewidth as well as
the frequency and linewidth of the cavity mode. Even though they do
not bump energy into the cavity but on the contrary absorb energy
from the cavity, the atoms as an active component strikes a gain-loss
balance to the rest two components. The gain-loss balance has recently
proven to be central to the study of coupled optomechanical ring cavities~\cite{jing14,bender13,peng14},
which can not only initiate phonon lasing~\cite{jing14} but also
trigger a chaotic motion~\cite{xylv15}.

In Sec.~\ref{sec:Hystersis}, we show the hystersis cycles, which
are prevalent in multistable cavity systems, peculiar to the \emph{$\mathcal{PT}$}-asymmetric
AOM system and discuss their properties. The conclusions are given
finally in Sec.~\ref{sec:Conclusions}.

\section{Coupling mechanism\label{sec:Coupling-mechanism} }

\begin{figure}
\includegraphics[clip,width=8cm]{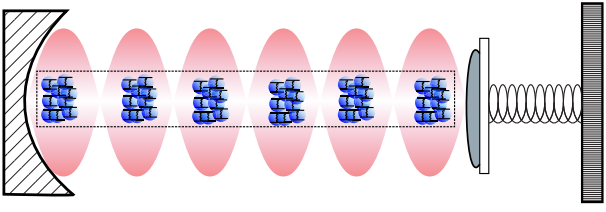}

\caption{(Color online) Illustration of the tripartite optomechanical cavity
with the active parts colored: each blue disk is a two-level atom
of uniform eigenenergy $\Omega$ that couples to a the red-shaded
single-mode cavity field of frequency $\omega_{C}$. The dark-gray
tinted cap is the oscillating mirror of frequency $\omega_{M}$ that
interacts through radiation pressure on the reflective mirror coating
at the right cavity wall.~\label{fig:model}}
\end{figure}

We consider the model illustrated in Fig.~\ref{fig:model}. A mirror
of effective mass $m$, whose displacement from the equilibrium position
is denoted by $x$ and conjugate momentum denoted by $p$, is undergoing
an oscillating motion of frequency $\omega_{M}$. The motions of the
mirror and the atoms are coupled via the cavity field of frequency
$\omega_{0}$: on one side, the cavity field acts on the mirror through
the radiation pressure $\eta$ initiated by the small displacement
$x$ of the mirror that deforms the cavity volume $V$; on the other,
the cavity field couples with the two-level atoms through dipole-field
interaction whose strength $g_{j}=\Omega\mu\sin k_{j}Q_{j}/\sqrt{(\omega_{0}+\eta x)\epsilon_{0}V(1+x)}$
with $\hbar=1$ is parametrized by the cavity geometry, where $Q_{j}$
denotes the position, $k_{j}$ the wave number, $\Omega$ the uniform
atomic level spacing, $\omega_{0}$ the cavity mode frequency, and
$\mu$ the uniform dipole moment of each atom.

For the densely packed atoms we consider here, particularly when they
form a Bose-Einstein condensate~\cite{brennecke07}, the uniformity
of the strength $g_{j}$ renders the atomic ensemble a Hopfield quantum
dielectric~\cite{hopfield58,ian08}. At low-energy, the atoms are
sparsely excited and these excitations can be collectively described
by a single bosonic mode $b$ that shares the same eigenenergy $\Omega$
with the $N$ individual atoms. This model leads the atom-field coupling
rescaled to 
\begin{equation}
G=\sqrt{n}g=\sqrt{\frac{N}{2\omega_{0}\epsilon_{0}V}}\Omega\mu\label{eq:Hopfield_coupling}
\end{equation}
where $n=N/V$ denotes the density of packed atoms. In other words,
when the scale of the atom-field detuning and the deformation of the
cavity volume caused by the displacement $x$ is relatively small,
the radiation pressure has no direct effect on the distribution of
the atoms.

Nonetheless, the cavity field acts as an optical lattice that distributes
the atoms periodically along the cavity axis~\cite{colombe07,dong11}.
The atoms consequently forms a grating that diffracts the cavity field,
leading to an optical gain~\cite{hemmer96}. To be exact, the cavity
field acts simultaneously as a quasiresonant pump field to the atoms,
which produces collective atomic recoil lasing without population
inversion~\cite{bonifacio94}. The recoil lasing is reflected by
a gain term with real positive coefficient in the atom-cavity coupled
equations of motion~\cite{javaloyes08}.

To observe more directly the relation between the dynamic gain-loss
and $\mathcal{PT}$-symmetry, we introduce a hypothetical ``phase
factor'' $e^{i\phi}$ in the tripartite AOM system Hamiltonian
\begin{eqnarray}
H & = & \Omega b^{\dagger}b+\omega_{0}a^{\dagger}a+\frac{p^{2}}{2m}+\frac{1}{2}m\omega_{M}^{2}x^{2}\nonumber \\
 &  & -\eta xa^{\dagger}a+Ge^{i\phi}(b^{\dagger}a+ba^{\dagger}).\label{eq:Ham}
\end{eqnarray}
The real part of $Ge^{i\phi}$ would then associate with the usual
dispersion while its imaginary part associates with an atom-cavity
gain in the corresponding equations of motion. Writing the two fast-varying
modes under the rotating frame of the cavity, i.e. $a\to ae^{-i\omega_{0}t}$
and $b\to be^{-i\omega_{0}t}$, these equations of motion read 
\begin{eqnarray}
\ddot{x} & + & \Gamma\dot{x}+\omega_{M}^{2}x=\frac{\eta}{m}a^{\dagger}a,\label{eq:EoM_x}\\
\dot{a} & = & i\eta xa-\kappa a-iGe^{i\phi}b,\label{eq:EoM_a}\\
\dot{b} & = & -i\delta b-\gamma b-iGe^{i\phi}a.\label{eq:EoM_b}
\end{eqnarray}
where $\delta=\Omega-\omega_{0}$ denotes the atom-cavity detuning,
$\Gamma$ the decay rate of the mirror, $\kappa$ the linewidth of
the cavity field, and $\gamma$ the relaxation rate of the bosonic
mode of the atoms. Decoupling Eqs.~\eqref{eq:EoM_a}-\eqref{eq:EoM_b},
we note that the ``phase'' $\phi$ taking values away from the period
$2n\pi$ would lead to a gain to the cavity mode $a$ when breaking
the $\mathcal{PT}$-symmetry of the Hamiltonian Eq.~\eqref{eq:Ham}
about the steady state $\left\langle x\right\rangle =\eta|\left\langle a\right\rangle |^{2}/m\omega_{M}^{2}$.

\section{Dynamic multistability\label{sec:Phase-matching}}

\subsection{Stability branching of mirror}

To see $\left\langle x\right\rangle $ can admit non-zero values,
we combine the steady-state solutions of Eqs.~\eqref{eq:EoM_x}-\eqref{eq:EoM_b}
to arrive at a gain-loss balance equation
\begin{equation}
\kappa\gamma+\delta\eta\left\langle x\right\rangle +G^{2}e^{2i\phi}+i\left[\kappa\delta-\gamma\eta\left\langle x\right\rangle \right]=0\label{eq:balance_eq}
\end{equation}
Introducing the compound gain-loss ratio
\begin{equation}
\rho=\frac{G^{2}}{\kappa^{2}}\cdot\frac{G^{2}}{\gamma^{2}+\delta^{2}}\label{eq:gain_loss_ratio}
\end{equation}
of the AOM system, we derive from the balance equation that only when
the atom-cavity phase $\phi$ matches a periodic value
\begin{equation}
\phi_{0}=\frac{1}{2}\left[\tan^{-1}\frac{\delta\pm\gamma\sqrt{\rho-1}}{\gamma\mp\delta\sqrt{\rho-1}}+k\pi\right]\label{eq:phi_0}
\end{equation}
where $k\in\mathbb{Z}$, then the mirror can admit two non-zero steady
states
\begin{equation}
\left\langle x\right\rangle =\pm\frac{\kappa}{\eta}\sqrt{\rho-1}.\label{eq:steady_x}
\end{equation}

Since $\Omega$ and $\gamma$ are always greater than zero, the fraction
in the arctangent of Eq.~\eqref{eq:phi_0} ensures that multiples
of $2\pi$ would be not admissible for $\phi_{0}$, showing that multistability
only appears when $\mathcal{PT}$-symmetry is broken. Further, the
factor $\sqrt{\rho-1}$ in Eq.~\eqref{eq:phi_0} and Eq.~\eqref{eq:steady_x}
implies that multistability is admissable only when the atom-cavity
interaction $G$ enters the strong-coupling regime to overcome the
losses due to the cavity leakage and the atomic relaxation and to
offset the frequency of the atomic mode. 

\begin{figure}
\includegraphics[bb=30bp 0bp 755bp 592bp,clip,width=8cm]{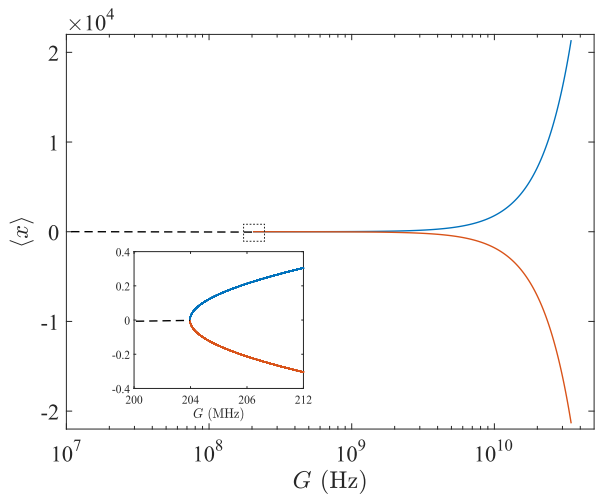}\caption{(Color online) Branching of the steady state $\left\langle x\right\rangle $
as a function of the coupling strength $G$ , plotted with abscissa
in log scale. The inset shows the magnified area in the dashed box
with abscissa in normal scale.~\label{fig:x_branching}}

\end{figure}

Figure~\ref{fig:x_branching} shows a semilog plot of $\left\langle x\right\rangle $
against $G$ over a wide range from $N=1$ to $N=10^{7}$, where the
parameters are set to experimentally accessible values~\cite{brennecke08}.
That is, $\kappa=1.3$~MHz, $\gamma=3.0$~MHz, $\eta=\sqrt{1.8}\kappa$,
$g=10.9$ MHz, and $\delta=32$~GHz. Setting off from $\left\langle x\right\rangle =0$,
the non-zero values admissable by $\left\langle x\right\rangle $
starts branching at $\rho=1$, which occurs at $G=204$~MHz in the
plot, into two branch cuts with different behaviors near and far from
the singular point. In the far right where $G\gg1$, $\left\langle x\right\rangle $
is quadratic whereas in the near range, $\left\langle x\right\rangle $
is parabolic. Specifically, seeing the two branches are monotonic
functions of $G$, we can define the saddle point where $d^{2}\left\langle x\right\rangle /dG^{2}=0$
at
\begin{equation}
G=\left[3\kappa^{2}(\gamma^{2}+\delta^{2})\right]^{1/4}
\end{equation}
to be the separating point between the quadratic region and the parabolic
region.

It should be remarked that, in classical steady-state models for FP
cavities, the intra-cavity atoms act as a dielectric that feeds a
nonlinear response to the cavity mode. Regarding the fields at the
two reflective mirrors as input and output ends, respectively, one
recognizes that the nonlinear response lets the fields traverse multiple
times in the dielectric medium that associates with a field-quadrature
dependent high-order susceptibililty. The input-output response equation
therefore translates to a high-order algebraic equation of the transmitted
field, where the first-order (linear) response corresponds to the
zero stable state and the higher-order responses correspond to the
non-zero stable states.

In steady-state models for optomechanical cavities, one has similar
high-order algebraic equations for multistability except its variable
becoming the coordinate of the mirror, where the cavity field can
be regarded as the input. In other words, the determination of $\left\langle x\right\rangle $
is established upon the determination of the steady states $\left\langle a\right\rangle $
of the cavity mode. This is equivalent to the tripartite model here
under Bogoliubov approximation for atomic condensates, i.e. $\left\langle b\right\rangle =\left\langle b^{\dagger}\right\rangle \to\sqrt{N}$
while the time-dependence only exists in fluctuating terms $\delta b(t)$
and $\delta b^{\dagger}(t)$. Given the approximation, Eq.~\eqref{eq:EoM_b}
can be ignored when considering the steady-state solutions and Eq.~\eqref{eq:EoM_a}
reduces the solution of $\left\langle a\right\rangle $ to the form
of a normal laser driving, $\mathcal{E}/(\kappa-i\eta\left\langle x\right\rangle )$,
where $\mathcal{E}=G\sqrt{N}$ and $\phi$ assumes the value of $\pi/2$.
Note that such a consideration only shows a formal resemblance of
the low-order response of an atomic condensate to an external driving.
It does not retrieve the steady state induced by external driving
since the limiting process does not recover the Hermitean and thus
$\mathcal{PT}$-symmetric interaction Hamiltonian $i(a-a^{\dagger})$
under the rotating frame. $\pi/2$ is not an admissible value given
by Eq.~\eqref{eq:phi_0}.

Henceforth, the atomic ensemble depicted collectively by the bosonic
mode does not play an active role in generating the nonlinear response
to obtain non-zero stable states to $\left\langle x\right\rangle $.
It instead feeds a complex response arising from its composite gain
and loss back to the cavity mode. This response cancels the degree
of freedom in the cavity mode exactly when the gain-loss balance is
obtained, i.e. when the phase angle $\phi$ matches $\phi_{0}$ given
in Eq.~\eqref{eq:phi_0}, where $\left\langle x\right\rangle $ breaks
off from the zero state along the two branches given in Eq.~\eqref{eq:steady_x}.
The steady states of $\left\langle x\right\rangle $ is thus determined
not by the cavity steady state $\left\langle a\right\rangle $, but
by the system parameters $\{G,\delta,\gamma,\kappa,\eta\}$. Rather,
the cavity steady state $\left\langle a\right\rangle $ is conversely
determined by $\left\langle x\right\rangle $.

Put in algebraic terms, the vector $\left\langle \mathbf{u}\right\rangle =(\left\langle a\right\rangle ,\left\langle b\right\rangle )$
of the steady states of the cavity mode and the atomic bosonic mode
is determined by Eqs.~. Written in matrix form, the system of equations
is the homogeneous equation 
\begin{equation}
\left[\begin{array}{cc}
\kappa-i\eta\left\langle x\right\rangle  & iGe^{i\phi}\\
iGe^{i\phi} & i\delta+\gamma
\end{array}\right]\left\langle \mathbf{u}\right\rangle =0,\label{eq:matrix_eq}
\end{equation}
which admits non-zero solutions only when the matrix is degenerate.
The branches plotted in Fig.~\ref{fig:x_branching} is the degenerate
curve of the matrix when the phase angle $\phi$ is fixed on one of
values admissible by Eq.~\eqref{eq:phi_0}.

\subsection{Phase-matching for gain-loss balance}

\begin{figure}
\includegraphics[bb=0bp 0bp 322bp 235bp,clip,width=7cm]{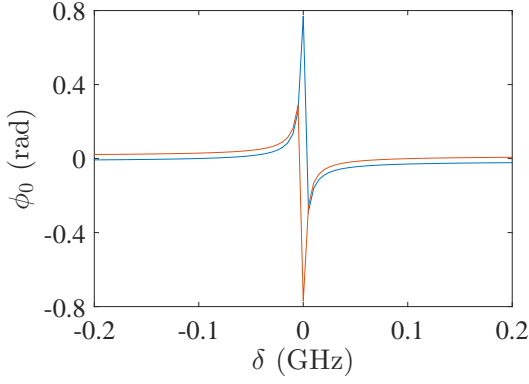}

\caption{(Color online) Matching phase angle $\phi_{0}$ versus the atom-cavity
detuning $\delta$ for the two branches of steadf states of $\left\langle x\right\rangle $.
The color of the curve matches that of its corresponding curve of
$\left\langle x\right\rangle $ in Fig.~\ref{fig:x_branching}. \label{fig:phi_0} }

\end{figure}

When $\left\langle x\right\rangle $ takes values along the branches
for any given atom-cavity coupling $G$, the phase angle $\phi$ has
to match with $\phi_{0}$ determined by the decay rates of the cavity
and the atoms and the detuning between them. This hypothetical angle
determines how the atom-cavity interaction breaks down into the real
and the imaginary parts and thus decides whether the atoms are in
the domain of gain or loss for the cavity field. Shown in Fig.~\ref{fig:phi_0}
with $k=0$, this phase-matching angle extends a wide range over the
detuning $\delta$, where the parameters are assumed the same values
taken by the plot in Fig.~\ref{fig:x_branching}. The two curves
correspond to the two diverging branches of the steady state $\left\langle x\right\rangle $
and they meet at two points $\delta=\pm\sqrt{(G^{4}/\kappa^{2})-\gamma^{2}}$
for which $\sqrt{\rho-1}=0$, which is the diverging point of $\left\langle x\right\rangle $. 

To determine whether the atom-cavity coupling is undergoing a gaining
process of a lossy processing for the cavity field at a given phase-matching
angle $\phi_{0}$, we examine the sign of the real part of the term
associated with the coefficient $iGe^{i\phi}$ in the equations of
motion. But seeing Eq.~\eqref{eq:EoM_a} and Eq.~\eqref{eq:EoM_b}
cannot be decoupled into a second-order differential equation of the
cavity mode $a$ alone, we determine the characteristics of the motion
of $a$ in the dual space of time through a Laplace transform.

Laplace transforming Eq.~\eqref{eq:EoM_b}, we find
\begin{equation}
\mathfrak{b}(s)=\frac{-iGe^{i\phi}\mathfrak{a}+\mathfrak{F}}{s+i\delta+\gamma},\label{eq:Lap_b}
\end{equation}
where $\mathfrak{a}$ and $\mathfrak{b}$ denote the cavity mode and
the atom bosonic mode, respectively, in their dual space. $\mathfrak{F}$
denotes the noise input associated with the dissipation term $\gamma b$.
Substituting Eq.~\eqref{eq:Lap_b} into the Laplace transform of
Eq.~\eqref{eq:EoM_a} and ignoring the noise input associated with
the cavity linewidth $\kappa$ as it is irrelevant to the cavity's
response to the motion of the atomic ensemble, we find the cavity
mode as the inverse Laplace transform
\begin{equation}
a(t)=\mathcal{L}^{-1}\left\{ \frac{-iGe^{i\phi}\mathfrak{F}}{(s-i\eta\mathfrak{x}+\kappa)(s+i\delta+\gamma)+G^{2}e^{2i\phi}}\right\} .\label{eq:inv_Lap}
\end{equation}

The denominator of Eq.~\eqref{eq:inv_Lap} is a quadratic equation
of $s$ whose two roots will become the coefficients of time in exponentials
through the inverse transform. Hence, the real parts of the roots
will signify the amplification or attenuation of the cavity mode while
the imaginary parts only contribute to the dispersed oscillations
of the cavity mode. For the atom-cavity coupling to set off the loss
in the cavity and the atomic modes and contribute a net gain to the
dynamic system, one requires from this quadratic equation that the
inequality
\begin{multline}
\sin^{2}2\phi+\frac{\kappa-\gamma}{G}\frac{\delta+\eta x}{G}\sin2\phi-\left(\frac{\kappa+\gamma}{G}\right)^{2}\cos2\phi\\
\geq\frac{\kappa\gamma}{G^{2}}\left[\left(\frac{\kappa+\gamma}{G}\right)^{2}+\left(\frac{\delta+\eta x}{G}\right)^{2}\right]\label{eq:inequality}
\end{multline}
regarding the phase factor $Ge^{i\phi}$ is satisfied (see Appendix
for the detailed derivation).

Shown in Fig.~\ref{fig:gain_loss_contour}, the contour plots illustrate
the inequality through the difference between the terms on the left
and on the right hand sides of Eq.~\eqref{eq:inequality} against
the variables $\delta$, $G$ and $\phi$, such that the contour levels
that are greater than zero (circled by the white dashed curves) indicate
the region where Eq.~\eqref{eq:inequality} is satisfied, i.e. where
the atom-cavity interaction obtains a net gain. The system parameters
used retain those used in Sec.~\ref{sec:Phase-matching}.A above.
In particular, the vertical axis for $G$ in the subfigure (a) corresponds
to the exponential scaling of experimentally accessible atomic density
from $10^{3}$ to $10^{6}$. For the subfigure (b), we choose $G$
to be 1GHz where the net gain regions just emerge.

\begin{figure}
\includegraphics[bb=20bp 0bp 475bp 322bp,clip,width=8cm]{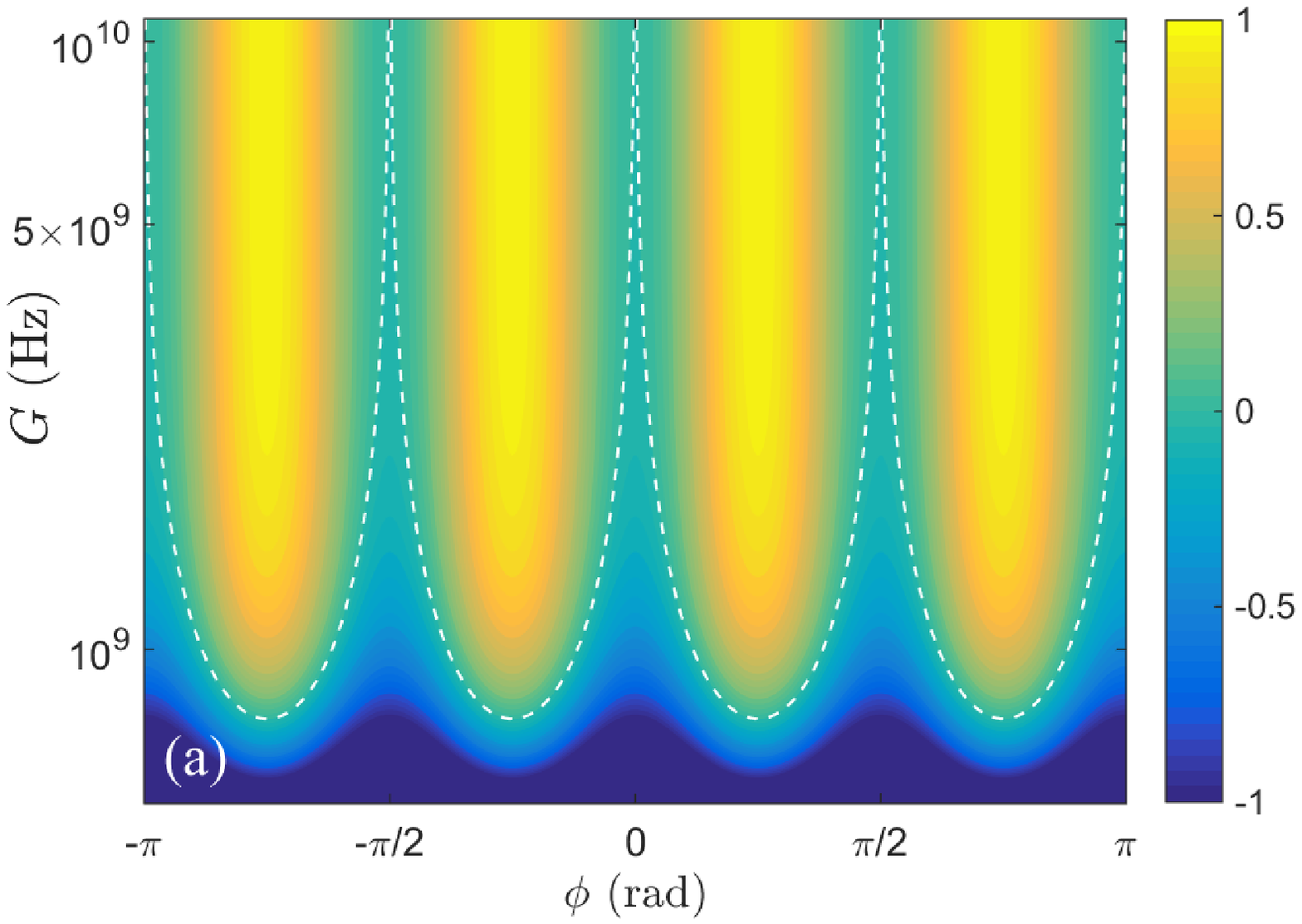}

\includegraphics[bb=5bp 6bp 442bp 300bp,clip,width=8cm]{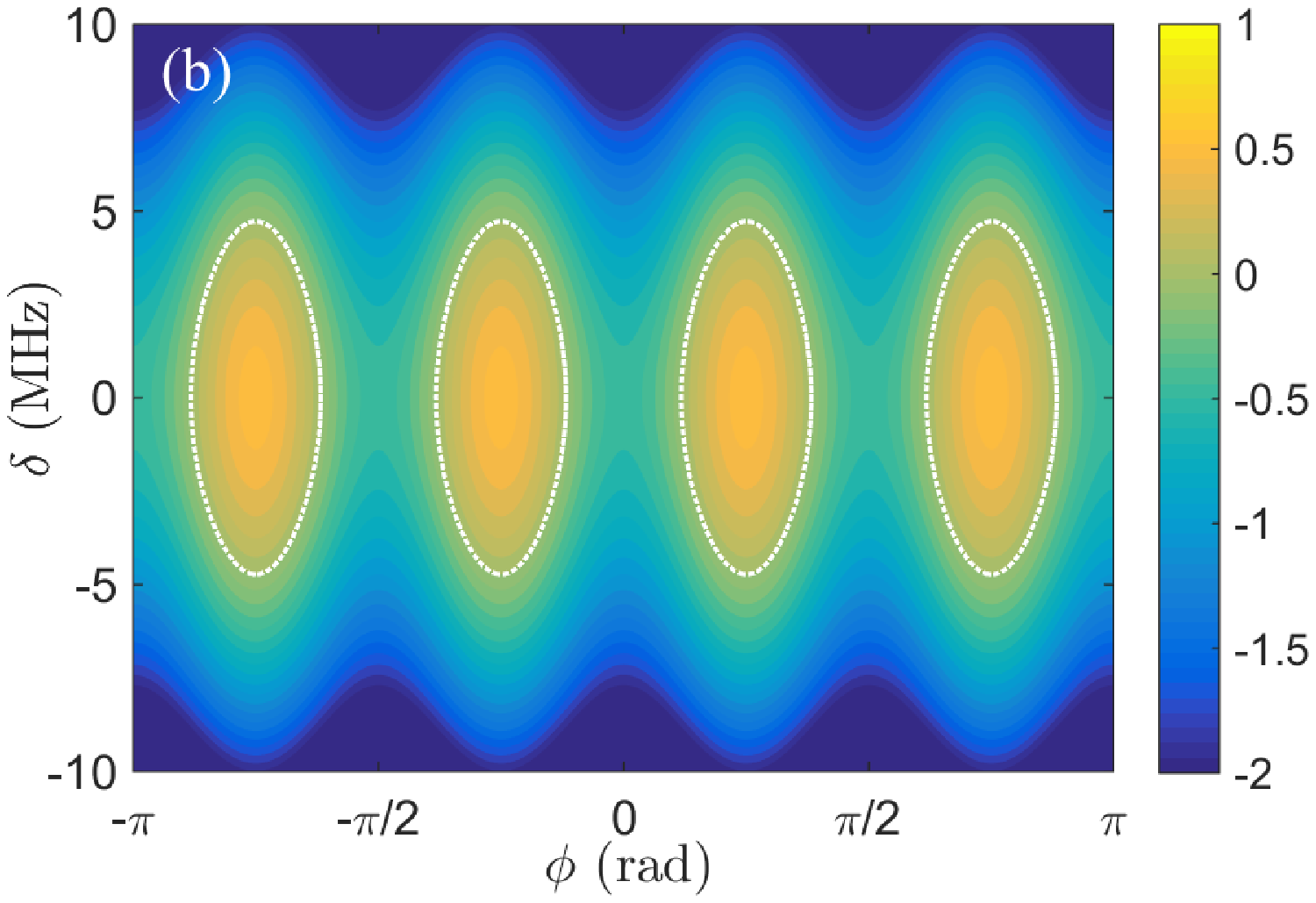}

\caption{(Color online) Illustrating the atom-cavity gain regions through contour
plots of the gain-loss inequality (a) as a function $F(G,\phi)$ against
the atom-cavity coupling strength $G$ and the phase $\phi$ at atom-cavity
resonance $\delta=0$; and (b) as a function $F(\delta,\phi)$ against
the atom-cavity detuning $\delta$ and the phase $\phi$ at $G=1$GHz.
In (a), the vertical $G$ axis is logarithmic to show the effect of
exponential scaling of atomic density $N$. In both plots, the white
dashed curves indicate the zero level, within which one obtains a
net gain and without which one obtains a net loss to the optomechanical
system.~\label{fig:gain_loss_contour}}
\end{figure}

The two plots here verify the analysis given in the last section that
multistable regions emerge when the atom-cavity coupling is strong
enough to overcome their individual dissipations to the environment.
The loss-to-gain transitions in the contours (the white dashed curves)
coincides with the phase-matching value of Eq.~\eqref{eq:phi_0}
for which the dynamic gain-loss balance between the atomic ensemble
and the cavity field as given in Eq.~\eqref{eq:balance_eq} is obtained.
The atoms therefore plays the active role under which the stable values
admissible to the mirror's coordinate $\left\langle x\right\rangle $
can be dynamically traced along the paths of Fig.~\ref{fig:x_branching}
through tuning the phase $\phi_{0}$ and the detuning $\delta$ along
a fixed contour in Fig.~\ref{fig:gain_loss_contour}.

Moreover, shown in subfigure (a), the net gain regions expand when
$G$ is increased and the regions become asymptotically connected
when $G$ approaches to infinitely large. Correspondingly in subfigure
(b), stronger coupling $G$ leads to both longer major and longer
minor axes in the elliptical net-gain regions. The strong-coupling
effect is more exemplified in the major axis, signifying a larger
tolerance range of gain about the detuning. No matter the net-gain
regions shrink or expand, their centers remain fixated at the atom-cavity
resonance on one axis and at the points where $\phi=[\pm\tan^{-1}(G^{2}/\kappa\gamma)+k\pi]/2$
on the other, the latter of which is reduced from Eq.~\eqref{eq:phi_0}
when $\delta$ is set to 0. Coinciding with the peaks of $\phi$ in
Fig.~\ref{fig:phi_0}, these periodic points in $\phi$ are invariable
with respect to $G$ and form the dynamical attractors of the mirror's
oscillation in the optomechanical cavity.

\section{Hystersis cycles\label{sec:Hystersis}}

When studying the multistability of a dielectric in an FP cavity,
the high-order susceptibility of the dielectric translates into a
multi-response of the classical medium to an input driving field.
This multi-response is reflected graphically as a hystersis cycle
in the input-output plot where the incident field strength or its
variants is regarded as the input and the transmitted field strength
or its variants as the output. Depending on the absorptive response
of the susceptibility, different shapes of hystersis cycles will result,
determining the existence of multistability and characterizing the
multistability if it does exist. For example, in one respect, the
absorptivity of the medium is reflected by the area enclosed in the
hystersis cycle.~\cite{Gibbs}

Here, we treat the coordinate quadrature $X_{\left\langle b\right\rangle }=\left\langle b\right\rangle +\left\langle b^{\dagger}\right\rangle $
of the steady state of the atomic ensemble as the driving input and
the coordinate quadrature $X_{\left\langle a\right\rangle }=\left\langle a\right\rangle +\left\langle a^{\dagger}\right\rangle $
of the cavity field as output to examine the hystersis cycle. From
Eq.~\eqref{eq:EoM_a}, the steady-state value $\left\langle a\right\rangle $
is proportional to $\left\langle b\right\rangle $. Conversing the
variable dependence and substituting the steady-state of the mirror
$\left\langle x\right\rangle $ from Eq.~\eqref{eq:EoM_x}, we have
\begin{equation}
\left\langle b\right\rangle =\frac{1}{Ge^{i\phi}}\left[\frac{\eta^{2}|\left\langle a\right\rangle |^{2}}{m\omega_{M}^{2}}+i\kappa\right]\left\langle a\right\rangle .
\end{equation}
Thus, besides a constant phase difference introduced by the proportional
coefficient, the variation of $\left\langle b\right\rangle $ is in-phase
with the variation of $\left\langle a\right\rangle $. 

The equations of motion Eq.~\eqref{eq:EoM_x}-~\eqref{eq:EoM_b}
do not determine this common phase shared by $\left\langle a\right\rangle $
and $\left\langle b\right\rangle $ since the steady states constrained
by the gain-loss balance in Eq.~\eqref{eq:balance_eq} make the matrix
equation in Eq.~\eqref{eq:matrix_eq} self-consistent, thereby leaving
an undetermined degree of freedom in this phase. To simplify our discussions
below, we choose zero for this phase such that $\left\langle a\right\rangle =\left\langle a^{\dagger}\right\rangle =|\left\langle a\right\rangle |$.
The quadrature of the atomic bosonic mode can hence be written as
\begin{equation}
X_{\left\langle b\right\rangle }=X_{\left\langle a\right\rangle }\left[\frac{\eta^{2}X_{\left\langle a\right\rangle }^{2}}{4Gm\omega_{M}^{2}}\cos\phi+\frac{\kappa}{G}\sin\phi\right],
\end{equation}
in a form reminiscent to the hystersis cycle of static bistability
for a dielectric in a FP cavity.

\begin{figure}
\includegraphics[bb=8bp 0bp 382bp 260bp,clip,width=8cm]{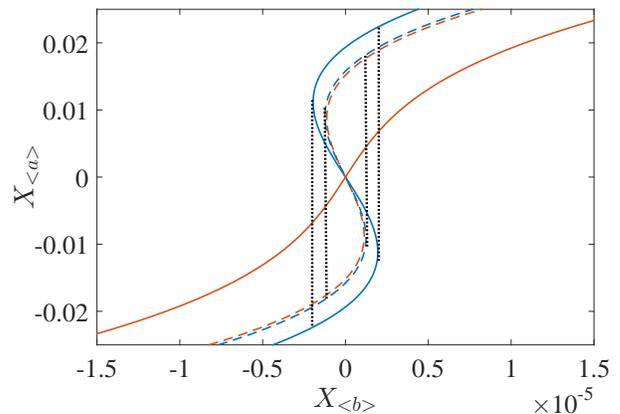}

\caption{(Color online) Hystersis cycles formed in the quadrature output $X_{\left\langle a\right\rangle }$
in the cavity field versus the quadrature input $X_{\left\langle b\right\rangle }$
in the atomic mode at atom-cavity resonance $\delta=0$ (solid curves)
and at small detuning $\delta=-1.5$MHz (dashed curves). The black
dotted lines serve as visual guides for the cycles formed. The blue
curves (solid and dashed) correspond to the upper branch of $\left\langle x\right\rangle $
while the orange curves (solid and dashed) correspond to the lower
branch of $\left\langle x\right\rangle $ in Fig.~\ref{fig:x_branching}.~\label{fig:hystersis}}
\end{figure}

However, the three-order polynomial on the right hand side of the
equation not only gives rise to a bi-valuation of $X_{\left\langle b\right\rangle }$
at a given $X_{\left\langle a\right\rangle }$ because the phase $\phi$
at the steady state of $\left\langle x\right\rangle $ can admit multiple
values constrained only by the coupling strength $G$ as given in
Eq.~\eqref{eq:phi_0}. Besides the periodicity allowed by the integer
values of $k$, $\phi_{0}$ can admit two values along the two branches
of $\left\langle x\right\rangle $ for a fixed $k$, therefore giving
rise to quadruple-valuation of $X_{\left\langle b\right\rangle }$
if $G$ is given appropriate values. This is shown in Fig.~\ref{fig:hystersis}
as two differently colored curves where $k=0$ and $G$ is chosen
to be $345$MHz. The detuning $\delta$ is chosen at two typical values:
$\delta=0$ and $\delta=-1.5$MHz. 

At resonance, the upper branch of $\left\langle x\right\rangle $
associates with two bistable values of $X_{\left\langle b\right\rangle }$
while the lower branch of $\left\langle x\right\rangle $, though
itself being mono-valued, adds a third valuation to $X_{\left\langle b\right\rangle }$,
making the system tristable. Departing from resonance, the values
of $X_{\left\langle b\right\rangle }$ associated with the two branches
of $\left\langle x\right\rangle $ converge, exhibiting bistability
for both branches, making the system quadruply stable, shown as the
dashed curves in the figure. We note that the asymmetric dependence
of $\phi$ on $\delta$ leads to an asymmetric dependence of the hystersis
cycles on $\delta$, i.e. the hystersis cycles vanish at $\delta=1.5$MHz.

\section{Conclusions\label{sec:Conclusions}}

Using a hypothetical Hamiltonian whose interaction part has broken
$\mathcal{PT}$-symmetry, we show that periodical multistable states
can exhibit in an atom-filled optomechanical cavity when the periodicity
matches a gain or loss balance between the filled atomic ensemble
and the cavity field. Unlike the multistability arises in FP cavity,
the atoms whose motion is described by a bosonic mode plays an active
role in developing the multistable states. When the gain brought by
the atoms into the cavity exactly counter-act against the loss of
all components in the cavity, the mechanical mirror is driven from
zero state into two branches that match with the amount of gain dynamically
obtained.

Since traditional studies on multistability strongly focus on the
controllability of optical systems, the dynamically tuned multistability
developed here can be viewed as an extra level of control one can
exert on optical cavity systems without the interference of an external
driving laser.

We also note that the collapse of the hystersis cycles, i.e. transition
from multi-valuation to mono-valuation of input-output, shown in Fig.~\ref{fig:hystersis}
is not continuous with respect to the detuning $\delta$ and we leave
the investigation of this discontinuity in transition to future studies.

\appendix

\section{Weighing gain through Laplace transforms}

The denominator in the inverse Laplace transform of Eq.~\eqref{eq:inv_Lap}
is the quadratic polynomial
\begin{multline}
s^{2}+\left[\kappa+\gamma+i(\delta-\eta\mathfrak{x})\right]s+\delta\eta\mathfrak{x}\\
+\kappa\gamma+i(\delta\kappa-\gamma\eta\mathfrak{x})+G^{2}e^{2i\phi}\label{eq:quad_root}
\end{multline}
whose zeros will concern the determination of gain and loss of the
cavity field contributed by the atom-cavity interaction. Since the
coefficient of the linear term of the polynomial is independent of
$G$ and $\phi$, we need only consider the discriminant of the roots.
Writing the phase $e^{2i\phi}$ in rectangular coordinate form and
combining it with the other terms, we find the complex discriminant
to be
\begin{multline*}
\mathcal{D}=(\kappa-\gamma)^{2}-(\delta+\eta\mathfrak{x})^{2}-4G^{2}\cos2\phi\\
-2i\left[(\kappa-\gamma)(\delta+\eta\mathfrak{x})+2G^{2}\sin2\phi\right].
\end{multline*}
Introducing the abbreviations $\kappa_{\gamma}=\kappa-\gamma$ and
$\delta_{\eta}=\delta+\eta\left\langle x\right\rangle $, it reads
in polar form
\begin{multline*}
\mathcal{D}=4G^{2}\exp\{-i\theta\}\times\\
\sqrt{1-\left[\frac{\kappa_{\gamma}^{2}-\delta_{\eta}^{2}}{2G^{2}}\cos2\phi-\frac{\kappa_{\gamma}\delta_{\eta}}{G^{2}}\sin2\phi\right]+\left(\frac{\kappa_{\gamma}^{2}+\delta_{\eta}^{2}}{4G^{2}}\right)^{2}}
\end{multline*}
where
\begin{equation}
\theta=\tan^{-1}\frac{2\kappa_{\gamma}\delta_{\eta}+4G^{2}\sin2\phi}{\kappa_{\gamma}^{2}-\delta_{\eta}^{2}-4G^{2}\cos2\phi}+k\pi\label{eq:theta}
\end{equation}
for all $k\in\mathbb{Z}$.

The sign of the real part of $\sqrt{\mathcal{D}}$, concerning the
amplification and attenuation, is determined by the sign of $\cos(\theta/2)$.
Considering that we let the arctangent take value between $-\pi/2$
and $\pi/2$ only, $\cos(\theta/2)$ would be positive and if $k$
is either 0 or even and negative if $k$ is odd. Therefore, as a part
of a root, the sign of $\sqrt{\mathcal{D}}$ is independent of the
values $G$ and $\phi$ take but is dependent on the sign of the square
root takes, with the positive corresponds to the upper branch and
the negative corresponds to the lower branch of Fig.~\ref{fig:x_branching}.

To decide whether the positive branch can contribute a net gain to
the atom-driven mirror oscillation, we compare its value to the decay
contributed by the cavity and the atoms themselves (the real part
of the linear term in Eq.~\eqref{eq:quad_root}), i.e.
\[
\Re\{\sqrt{\mathcal{D}}\}\geq\kappa+\gamma.
\]
Shuffling the terms, we arrive at the inequality
\begin{multline*}
\left(G^{2}\sin2\phi\right)^{2}+(\kappa-\gamma)\delta_{\eta}G^{2}\sin2\phi\\
-(\kappa+\gamma)^{2}G^{2}\cos2\phi\geq\kappa\gamma\left[(\kappa+\gamma)^{2}+\delta_{\eta}^{2}\right].
\end{multline*}
Dividing all the terms by $G^{4}$ gives Eq.~\eqref{eq:inequality}.
At the steady state, one has from the balance equation given in Eq.~\eqref{eq:balance_eq}
that
\begin{eqnarray*}
G^{2}\cos2\phi & = & -\kappa\gamma-\delta\eta\left\langle x\right\rangle ,\\
G^{2}\sin2\phi & = & -\kappa\delta+\gamma\eta\left\langle x\right\rangle ,
\end{eqnarray*}
for which one can verify that the equality sign is obtained, i.e.
$\Re\{\sqrt{\mathcal{D}}\}=\kappa+\gamma$. 
\begin{acknowledgments}
The research presented is supported by FDCT of Macau under grant 013/2013/A1,
University of Macau under grants MRG022/IH/2013/FST and MYRG2014-00052-FST,
and National Natural Science Foundation of China under grant No.~11404415.\end{acknowledgments}


\begin{thebibliography}{10}
\bibitem{gibbs76}H. M. Gibbs, S. L. McCall, and T. N. C. Venkatesan,
Phys. Rev. Lett. \textbf{36}, 1135 (1976); H. M. Gibbs, F. A. Hopf,
D. L. Kaplan, and R. L. Shoemaker, \emph{ibid}. \textbf{46}, 474 (1981).

\bibitem{Gibbs}H. Gibbs, \emph{Optical Bistability: Controlling Light
With Light} (Elsevier, 1985).

\bibitem{bonifacio78}R. Bonifacio and L. A. Lugiato, Phys. Rev. A
\textbf{18}, 1129 (1978); R. Bonifacio and P. Meystre, Optics Communications
\textbf{29}, 131 (1979).

\bibitem{marburger78}J. H. Marburger and F. S. Felber, Phys. Rev.
A \textbf{17}, 335 (1978).

\bibitem{kitano81}M. Kitano, T. Yabuzaki, and T. Ogawa, Phys. Rev.
Lett. \textbf{46}, 926 (1981).

\bibitem{joshi03}A. Joshi and M. Xiao, Phys. Rev. Lett. \textbf{91},
143904 (2003). 

\bibitem{meystre85}P. Meystre, E. M. Wright, J. D. McCullen, and
E. Vignes, J. Opt. Soc. Am. B \textbf{2}, 1830 (1985). 

\bibitem{gong09}Z. R. Gong, H. Ian, Y. Liu, C. P. Sun, and F. Nori,
Phys. Rev. A \textbf{80}, 065801 (2009).

\bibitem{klimov01}A. B. Klimov, L. L. Sánchez-Soto, and J. Delgado,
Optics Communications \textbf{191}, 419 (2001). 

\bibitem{dorsel83}A. Dorsel, J. D. McCullen, P. Meystre, E. Vignes,
and H. Walther, Phys. Rev. Lett. \textbf{51}, 1550 (1983).

\bibitem{marquardt06}F. Marquardt, J. G. E. Harris, and S. M. Girvin,
Phys. Rev. Lett. \textbf{96}, 103901 (2006).

\bibitem{chang11}Y. Chang, T. Shi, Y. Liu, C. P. Sun, and F. Nori,
Phys. Rev. A \textbf{83}, 063826 (2011).

\bibitem{marquardt09}F. Marquardt and S. M. Girvin, Physics \textbf{2},
40 (2009).

\bibitem{qian12}J. Qian, A. A. Clerk, K. Hammerer, and F. Marquardt,
Phys. Rev. Lett. \textbf{109}, 253601 (2012).

\bibitem{ghobadi11}R. Ghobadi, A. R. Bahrampour, and C. Simon, Phys.
Rev. A \textbf{84}, 033846 (2011).

\bibitem{ian15}T. Huan, R. Zhou, and H. Ian, Phys. Rev. A \textbf{92},
022301 (2015). 

\bibitem{dong11}Y. Dong, J. Ye, and H. Pu, Phys. Rev. A \textbf{83},
031608 (2011).

\bibitem{agarwal10}G. S. Agarwal and S. Huang, Phys. Rev. A \textbf{81},
041803 (2010).

\bibitem{kronwald13}A. Kronwald and F. Marquardt, Phys. Rev. Lett.
\textbf{111}, 133601 (2013).

\bibitem{karuza13}M. Karuza, C. Biancofiore, M. Bawaj, C. Molinelli,
M. Galassi, R. Natali, P. Tombesi, G. Di Giuseppe, and D. Vitali,
Phys. Rev. A \textbf{88}, 013804 (2013).

\bibitem{wallquist10}M. Wallquist, K. Hammerer, P. Zoller, C. Genes,
M. Ludwig, F. Marquardt, P. Treutlein, J. Ye, and H. J. Kimble, Phys.
Rev. A \textbf{81}, 023816 (2010).

\bibitem{hunger10}D. Hunger, S. Camerer, T. W. H\"{a}nsch, D. K\"{o}nig,
J. P. Kotthaus, J. Reichel, and P. Treutlein, Phys. Rev. Lett. \textbf{104},
143002 (2010).

\bibitem{zhang10}K. Zhang, W. Chen, M. Bhattacharya, and P. Meystre,
Phys. Rev. A \textbf{81}, 013802 (2010). 

\bibitem{brennecke08}F. Brennecke, S. Ritter, T. Donner, and T. Esslinger,
Science \textbf{322}, 235 (2008). 

\bibitem{ian08}H. Ian, Z. R. Gong, Y. Liu, C. P. Sun, and F. Nori,
Phys. Rev. A \textbf{78}, 013824 (2008).

\bibitem{brennecke07}F. Brennecke, T. Donner, S. Ritter, T. Bourdel,
M. Köhl, and T. Esslinger, Nature \textbf{450}, 268 (2007). 

\bibitem{hopfield58}J. J. Hopfield, Phys. Rev. \textbf{112}, 1555
(1958).

\bibitem{bender98}C. M. Bender and S. Boettcher, Phys. Rev. Lett.
\textbf{80}, 5243 (1998).

\bibitem{jing14}H. Jing, S. K. Özdemir, X.-Y. Lü, J. Zhang, L. Yang,
and F. Nori, Phys. Rev. Lett. \textbf{113}, 053604 (2014). 

\bibitem{bender13}C. M. Bender, M. Gianfreda, \c{S}. K. \"{O}zdemir,
B. Peng, and L. Yang, Phys. Rev. A \textbf{88}, 062111 (2013).

\bibitem{peng14}B. Peng, \c{S}. K. \"{O}zdemir, F. Lei, F. Monifi,
M. Gianfreda, G. L. Long, S. Fan, F. Nori, C. M. Bender, and L. Yang,
Nat. Phys. \textbf{10}, 394 (2014).

\bibitem{xylv15}X.-Y. L\"{u}, H. Jing, J.-Y. Ma, and Y. Wu, Phys.
Rev. Lett. \textbf{114}, 253601 (2015).

\bibitem{colombe07}Y. Colombe, T. Steinmetz, G. Dubois, F. Linke,
D. Hunger, and J. Reichel, Nature \textbf{450}, 272 (2007). 

\bibitem{hemmer96}P. R. Hemmer, N. P. Bigelow, D. P. Katz, M. S.
Shahriar, L. DeSalvo, and R. Bonifacio, Phys. Rev. Lett. \textbf{77},
1468 (1996).

\bibitem{bonifacio94}R. Bonifacio, L. De Salvo, L. M. Narducci, and
E. J. D\textquoteright Angelo, Phys. Rev. A \textbf{50}, 1716 (1994).

\bibitem{javaloyes08}J. Javaloyes, M. Perrin, and A. Politi, Phys.
Rev. E \textbf{78}, 011108 (2008). \end{thebibliography}
\end{document}